\documentstyle[amstex,aps,epsfig]{revtex}
\begin{document}
\title{\hfill TPJU-10/2001\\
~~ \\
QCD condensates and the pion wave functions in the nonlocal chiral model}
\author{{\bf Micha{\l} Prasza\l{}owicz and Andrzej Rostworowski}}
\address{{\em M.Smoluchowski Institute of Physics,} \\
{\em Jagellonian University,} \\
{\em Reymonta 4, 30-059 Krak\'{o}w, Poland.}}
\maketitle

\begin{abstract}
We use the simple instanton motivated Nambu Jona-Lasinio - type model to
calculate a twist 3 pseudo-scalar pion light cone wave function
$\psi_{\pi}^{PS}$. Using the normalization condition for $\psi_{\pi}^{PS}$
we calculate the quark condensate and also the gluon condensate, which agree
with the phenomenological values for these quantities. Since we can compute
also the $k_T^2$ dependence of the light cone wave functions, we calculate
$k_T^2$ moments of the pseudo-scalar and axial-vector wave functions which
are related to the mixed vacuum condensates. This allows us to extract the
condensates and compare them with existing estimates.
\end{abstract}


\section{Introduction}

\label{s:intro}

Soft pion theorems provide link between dynamical objects like the light
cone wave functions \cite{ChZh,BrLep,FarJack,EfRad} and static properties of
the physical vacuum \cite{ChZhitZhit,Zhitkt4} (for review see \cite
{Zhitnitsky:1997mc}). Customarily in QCD pion properties are expressed in
terms of the nonperturbative condensates, which are estimated by some
theoretical model, like the QCD sum rules \cite{SVZ} (see for example \cite
{ChZhPRep}), or the instanton model of the QCD vacuum \cite
{Polyakov:1996kh,Polyakov:1998ip} (for recent review see {\em e.g.} \cite
{DPrev}). Here we propose to invert this logic. In Ref.\cite
{Praszalowicz:2001wy} we have discussed a simple and tractable model, which
describes very well the leading twist pion wave function $\psi_{\pi}^{AV}$
not only in longitudinal variable $u$, but also in $k_{T}$. In this paper we
calculate the twist 3 pion wave function $\psi_{\pi}^{PS}$ \cite
{ChZhPRep,BrauFil,BrauFil2,Ball:1999je} and try to extract from both $%
\psi_{\pi}^{AV}$ and $\psi_{\pi}^{PS}$ the quark condensate and
mixed condensates of dimension 5 and 7, and the gluon condensate
as well. The model, which we use is a simple nonlocal chiral
pion-quark model \cite {PetPob,Bochum}, which in principle can be
obtained from the instanton model of the QCD vacuum \cite{DPrev}.
However, since we want to work directly in the Minkowski space, we
have to modify the form of the nonlocality \cite
{Praszalowicz:2001wy,PetPob,Bochum} in order to make the
calculations feasible. The price is that, of course, we loose the
direct contact with the original model. Therefore our aim is
twofold. Apart from giving estimates of the condensates we want to
perform tests in order to gain confidence in the model as well as
to find its limitations.

In most applications of the effective quark-meson theory described above, an
approximation of the constant constituent quark mass was used. This approach
has an impressive success in describing baryon (which emerge as solitons)
properties (see \cite{DPrev} and references therein). Recently an attempt to
include the momentum dependence in calculating solitons in the nonlocal NJL
model has been reported in Refs.\cite{BroRipGol}.

The instanton model of the QCD vacuum is formulated in the Euclidean
space-time, whereas the light cone wave functions are naturally defined
in the Minkowski space-time. One can in principle Wick rotate the integrals
and perform the calculations in the Euclidean metric, as has been for
example done in Refs.\cite{ET,DT,ADT} (for early attempts to calculate the
pion wave function in the instanton model see Ref. \cite{Ryskin}), however
in this work we shall perform all calculations directly in the Minkowski
space-time.

The instantons induce both gluon condensate, which is proportional
to the density of instantons, and quark condensate, which emerges due
to the delocalization of the quark zero modes \cite{DPrev,DP}. Therefore
both quark and gluon condensation occurs at the same scale $Q_0$, which
is associated with the average instanton size $1/\overline{\rho} = 600$~MeV.
After integrating out gluons and performing the bosonization, the model
reduces to the simple Nambu-Jona Lasinio type model where the quarks interact
{\em nonlocally} with an external meson field $U$ \cite{DPrev,DP}:
\begin{equation}
S_{I}=\int \frac{d^{4}kd^{4}l}{(2\pi )^{8}}\bar{\psi}(k)\sqrt{M(k)}U^{\gamma
_{5}}(k-l)\sqrt{M(l)}\psi (l)\,  \label{SI}
\end{equation}
and $U^{\gamma _{5}}$ can be expanded in terms of the pion fields:
\begin{equation}
U^{\gamma _{5}}=1+\frac{i}{F_{\pi }}\gamma ^{5}\tau ^{A}\pi ^{A}-\frac{1}{%
2F_{\pi }^{2}}\pi ^{A}\pi ^{A}+\ldots  \label{U}
\end{equation}
Here $F_{\pi }=93$ MeV and $M(k)$ is a momentum dependent constituent quark
mass. In principle $M(k)$ has been calculated in the instanton model in
Euclidean space time. Here, following Refs.\cite{PetPob,Bochum}, we wish to
perform calculations directly in the Minkowski space. To this end we shall
choose a simple pole formula \cite{Praszalowicz:2001wy}
\begin{equation}
M(k)=M\left( -\frac{\Lambda ^{2}}{k^{2}-\Lambda ^{2}+i\epsilon }\right) ^{2n}
\label{cutform}
\end{equation}
which for $n\sim 2-3$ and for $k^{2}<0$ reproduces the $k$ dependence
obtained from the instantons reasonably well \cite{Praszalowicz:2001wy}. In
order to check model sensitivity to the specific form of the cutoff function
(\ref{cutform}) we shall vary $M$ from $325$ to $400$ MeV and $n$ from $1$
to $5$. Notice that $M(k)$ of Eq.(\ref{cutform}) provides an UV cutoff
for the loop integrals entering the expressions for the pion wave functions.
This is true both in Euclidean and in Minkowski space.

Using this simple prescription we  study various condensates, which are
related to the normalization and $k_{T}^{2}$ moments of leading twist
axial-vector (AV) and nonleading twist pseudo-scalar (PS) pion light cone
wave functions (for definitions see {\em e.g.} Ref.\cite
{BrauFil2,Ball:1999je}) as well as the PS wave function itself.

>From the point of view of QCD the quantities we calculate depend on a
nonperturbative scale $Q_0$ which, however, must not be confused neither
with the constituent mass $M$ nor with an auxiliary parameter $\Lambda$,
which is fixed by the normalization condition for $\psi_{\pi}^{AV}$.
For $k^2<0$ the Ansatz (\ref{cutform}) should imitate $M(k)$ obtained
from the instantons. And for the latter, as
explained above,  $Q_0 \sim 1/\overline{\rho} = 600$~MeV. It is therefore
natural to assume that $Q_0$ is of the order of a few hundred MeV
irrespectively of $M$ and $\Lambda$. The precise definition of $Q_0$ is
only possible within QCD and in all  effective models one can use only
qualitative {\em order of magnitude} arguments to estimate $Q_0$. A more
practical way to determine $Q_0$ will be discussed in Sect.\ref{s:kt}
where we will associate $Q_0$ with transverse integration cutoff $K_T$
which, as we shall see, is of the order of $760 < K_T < 1100$~MeV.

As a first step we calculate the quark condensate, which enters the
normalization of the PS pion light cone wave function. The numerical result
depends upon $M$ and $n$ and varies between $-(258$ Mev$)^{3}$ and $-(328$
MeV$)^{3}$. The commonly used phenomenological value of $\left\langle \bar{q}%
q\right\rangle $ is lower and reads approximately $-(250$ MeV$)^{3}$.
Therefore our highest value for the quark condensate overshoots the
phenomenological value by a factor of $2$. The reason for this rather too
high values of $\left\langle \bar{q}q\right\rangle $ is a relatively slow
convergence of the $dk_{T}^{2}$ integral for this quantity. This is even
more visible in the case of the $k_{T}^{2}$ moments of the PS wave function.
Indeed, one can relate both $\left\langle k_{T}^{2}\right\rangle _{AV}$ and $%
\left\langle k_{T}^{2}\right\rangle _{PS}$ to the same mixed condensate of
dimension $5$, $\left\langle ig\,\bar{q}\,\sigma \cdot G\,q\right\rangle $,
which yields $\left\langle k_{T}^{2}\right\rangle _{AV}/\left\langle
k_{T}^{2}\right\rangle _{PS}=5/9$ \cite{ChZhitZhit}. In our case this ratio
is two times smaller. It is, however, interesting to observe, that if we cut
the $dk_{T}^{2}$ at some $(k_{T}^{2})_{\max }=K_{T}^{2}$, which we choose in
such a way that $\left\langle \bar{q}q\right\rangle =-(250$ MeV$)^{3}$, then
the ratio of $\left\langle k_{T}^{2}\right\rangle $ is stable and
approximately equal to $5/9$. The mixed condensate $\left\langle ig\,\bar{q}%
\,\sigma \cdot G\,q\right\rangle $ comes out between $-(427$ MeV$)^{5}$ and $%
-(443$ MeV$)^{5}$ in a fair agreement with the estimates from the QCD sum
rules Ref.\cite{BelIof}. Unfortunately the higher moments of the PS wave
function are still too large even with a $k_{T}^{2}$ cutoff.

One of the advantages of our method is that the analytical expression for
the quark condensate is given in terms of a Minkowskian integral, which in a
limit of a constant $M(k)$ and $k_{M}^{2}\rightarrow-k_{E}^{2}$ reduces to
the well known Euclidean form. By comparing the two expressions one can by
inspection guess a continuation prescription, which allows to rewrite certain
Euclidean integrals as the Minkowskian ones. We use this, in some respect
{\em ad hoc}, prescription to calculate the gluon condensate with very
encouraging result: $\left\langle \alpha/\pi\,GG\right\rangle =(378$ MeV$%
)^{4}$ to $(383$ MeV$)^{4}$.

Finally we plot the PS pion light cone wave function for our set of model
parameters. We find that $\phi _{\pi }^{PS}(u)$ vanishes at the end points $%
u=0$ and $u=1$ except for the case of $n=1$. The latter behavior is in
agreement with the results obtained in Ref.\cite{Ball:1999je} within the QCD
sum rules. It is interesting to observe, that vanishing of $\phi _{\pi
}^{PS}(u)$ at the end points is correlated with the fact that $\phi _{\pi
}^{AV}(u)$ is concave at $u\sim 1$ and $u\sim 0$ for $n>1$.

The paper is organized as follows: in Section \ref{s:cond} we shortly
summarize the results of Refs.\cite{ChZhitZhit,Zhitkt4} concerning the
relation of the $k_{T}^{2}$ moments of the pion wave functions to the mixed
quark-gluon condensates. In Sect. \ref{s:piwfMk}, following Ref.\cite
{Praszalowicz:2001wy}, we explain how to evaluate the loop integrals with
the momentum dependent quark mass $M(k)$ given by Eq.(\ref{cutform}). In
Sect. \ref{s:GG} we construct the continuation prescription discussed above
and calculate the gluon condensate. In Sect. \ref{s:num} we give the
numerical results and Sect. \ref{s:sum} contains summary and conclusions.
Some of the results presented here have been partially reported in
Ref.\cite{Dubr}.

\section{Pion wave functions and their relation to the vacuum condensates}

\label{s:cond}

We shall be dealing with the leading twist axial-vector (AV) and twist 3
pseudo-scalar (PS) wave functions defined as follows \cite
{BrauFil2,Ball:1999je}:
\begin{align}
\phi_{\pi}^{AV}(u) & =\frac{1}{i\sqrt{2}F_{\pi}}\int\limits_{-\infty
}^{\infty}\frac{d\tau}{\pi}e^{-i\tau(2u-1)(nP)}\left\langle 0\right| \bar{%
\psi}(n\tau)\rlap{/}n\gamma_{5}\psi\left( -n\tau\right) \left| \pi
^{+}(P)\right\rangle ,  \nonumber \\
\phi_{\pi}^{PS}(u) & =-(nP)\frac{F_{\pi}}{\sqrt{2}\left\langle \bar {q}%
q\right\rangle }\int\limits_{-\infty}^{\infty}\frac{d\tau}{\pi}%
e^{-i\tau(2u-1)(nP)}\left\langle 0\right| \bar{\psi}(n\tau)i\gamma_{5}\psi%
\left( -n\tau\right) \left| \pi^{+}(P)\right\rangle .  \label{Phidefs}
\end{align}
where we have chosen $n=(1,0,0,-1)$ as a light-cone vector parallel to $%
z_{\mu}=\tau n_{\mu}$ and $\tilde{n}=(1,0,0,1)$ parallel to $P_{\mu}$. For
any 4-vector $v$ we have:
\begin{equation}
v^{+}=n\cdot v,\quad v^{-}=\tilde{n}\cdot v,\quad v_{\mu}=\frac{v^{+}}{2}%
\tilde{n}_{\mu}+\frac{v^{-}}{2}n_{\mu}+v_{\mu}^{\bot}.
\end{equation}
Both wave functions are normalized to 1. The normalization condition for $%
\phi_{\pi}^{PS}$ yields therefore the expression for $\left\langle \bar {q}%
q\right\rangle $. Evaluation of the norm of $\phi_{\pi}^{PS}$ in the present
model gives:
\begin{equation}
\left\langle \bar{q}q\right\rangle =-4iN_{c}\int\frac{d^{4}k}{\left(
2\pi\right) ^{4}}\sqrt{M(k)M(k-P)}\frac{\,k(k-P)-M(k)M(k-P)}{%
(k^{2}-M^{2}(k))((k-P)^{2}-M^{2}(k-P))}.  \label{qbarqM}
\end{equation}
For $P\rightarrow0$ in the local case $M(k)=M$ and in Euclidean metric Eq.(%
\ref{qbarqM}) reduces to
\begin{equation}
\left\langle \bar{q}q\right\rangle =-i4N_{c}\int\frac{d^{4}k}{\left(
2\pi\right) ^{4}}\frac{M}{k^{2}-M^{2}}=-4N_{c}\int\frac{d^{4}k_{E}}{\left(
2\pi\right) ^{4}}\frac{M}{k_{E}^{2}+M^{2}}.  \label{qbarqE}
\end{equation}
The last expression in (\ref{qbarqE}) is well defined only in Euclidean
metric and Eq.(\ref{qbarqM}) may serve as a prescription how it should be
continued to the Minkowski space. In fact, as we shall explicitly see, $%
\left\langle \bar{q}q\right\rangle $ of Eq.(\ref{qbarqM}) does not depend on
$P$.

Our approach enables to calculate not only the distributions in $u$ but in
fact full wave functions $\psi_{\pi}$ both in $u$ and $k_{T}^{2}$:
\begin{equation}
\phi_{\pi}(u)=\int\limits_{0}^{\infty}dk_{T}^{2}\,\psi_{\pi}(u,k_{T}^{2})
\end{equation}
both for AV and PS channels. Integrating first over $u$ gives the $k_{T}^{2}$
distribution
\begin{equation}
\tilde{\phi}_{\pi}(k_{T}^{2})=\int\limits_{0}^{1}du\,\psi_{\pi}(u,k_{T}^{2}).
\end{equation}

It has been shown in Refs.\cite{ChZhitZhit,Zhitkt4} that moments of $\tilde{%
\phi}_{\pi}(k_{T}^{2})$ are given in terms of the mixed quark-gluon
condensates
\begin{equation}
\left\langle k_{T}^{2}\right\rangle _{AV}=\frac{5}{36}\frac{\left\langle ig\,%
\bar{q}\,\sigma\cdot G\,q\right\rangle }{\left\langle \bar{q}q\right\rangle }%
,\;\left\langle k_{T}^{2}\right\rangle _{PS}=\frac{1}{4}\frac{\left\langle
ig\,\bar{q}\,\sigma\cdot G\,q\right\rangle }{\left\langle \bar{q}%
q\right\rangle }.  \label{kt2}
\end{equation}
Here $G_{\mu\nu}^{a}$ is a gluon field strength and
\begin{equation}
\sigma\cdot G=\sigma_{\mu\nu}G^{\mu\nu},\;G_{\mu\nu}=\;\frac{\lambda^{a}}{2}%
G_{\mu\nu}^{a}.
\end{equation}
Note that formula for $\left\langle k_{T}^{2}\right\rangle _{AV}$ is an
approximate one since the soft pion theorems used to derive (\ref{kt2})
apply strictly only in the pseudo scalar channel. Equations (\ref{kt2})
predict
\begin{equation}
\left\langle k_{T}^{2}\right\rangle _{AV}=\frac{5}{9}\left\langle
k_{T}^{2}\right\rangle _{PS}.  \label{ratio}
\end{equation}
For $\left\langle k_{T}^{4}\right\rangle $ we have \cite{Zhitkt4}:
\begin{align}
\left\langle k_{T}^{4}\right\rangle _{AV} & =-\frac{3}{32}\frac{\left\langle
g^{2}\,\bar{\psi}\sigma\cdot G\,\sigma\cdot G\psi\right\rangle }{%
\left\langle \bar{\psi}\psi\right\rangle }+\frac{13}{72}\frac{\left\langle
g^{2}\,\bar {\psi}G^{2}\psi\right\rangle }{\left\langle \bar{\psi}%
\psi\right\rangle },  \nonumber \\
\left\langle k_{T}^{4}\right\rangle _{PS} & =-\frac{1}{12}\frac{\left\langle
g^{2}\,\bar{\psi}\sigma\cdot G\,\sigma\cdot G\psi\right\rangle }{%
\left\langle \bar{\psi}\psi\right\rangle }+\frac{1}{6}\frac{\left\langle
g^{2}\,\bar{\psi }G^{2}\psi\right\rangle }{\left\langle \bar{\psi}%
\psi\right\rangle }  \label{kt4}
\end{align}
where for $\left\langle k_{T}^{4}\right\rangle _{AV}$ the same
reservations as for $\left\langle k_{T}^{2}\right\rangle _{AV}$
hold. Unfortunately equations (\ref{kt4}) are almost linearly
dependent. Indeed, coefficients in front of the condensates are of
the order of $1/10$, whereas the determinant of (\ref{kt4}) is of
the order $6\times10^{-4}$. Therefore it is difficult to
disentangle the condensates from the knowledge of $\left\langle
k_{T}^{4}\right\rangle $, unless the latter are known with very
high accuracy.

\section{Pion wave function with momentum dependent constituent quark mass}

\label{s:piwfMk}

In our previous paper \cite{Praszalowicz:2001wy} we have shown how to handle
the loop integral, like the one in Eq.(\ref{qbarqM}), with momentum
dependent mass $M(k)$ given by Eq.(\ref{cutform}). To this end one
introduces the light cone integration variables
\begin{equation}
d^{4}k=\frac{1}{2}dk^{+}dk^{-}d^{2}\vec{k}_{T},\qquad k^{+}=uP^{+}.
\end{equation}
The key point of this analysis consists in finding the integration contour
in the complex $k^{-}$ plane. The poles of the fermion propagator are given
by the zeros of the following function:
\[
k^{2}-M^{2}\left[ \frac{\Lambda^{2}}{k^{2}-\Lambda^{2}+i\epsilon}\right]
^{4n}+i\epsilon=0
\]
which after pulling out the factor $\Lambda^{2}\left[ k^{2}/\Lambda
^{2}-1+i\epsilon\right] ^{-4n}$ reduces to
\begin{equation}
\left( \frac{k^{2}}{\Lambda^{2}}-1+i\epsilon\right) ^{4n+1}+\left( \frac{%
k^{2}}{\Lambda^{2}}-1+i\epsilon\right) ^{4n}-\frac{M^{2}}{\Lambda^{2}}=0
\label{denom}
\end{equation}
and similarly for $k\rightarrow k-P$. Here
\begin{equation}
k^{2}=uP^{+}k^{-}-\vec{k}_{T}^{\,2}\quad\text{and}%
\quad(k-P)^{2}=-(1-u)P^{+}k^{-}-\vec{k}_{T}^{\,2}
\end{equation}
Eq.(\ref{denom}) should be understood as an equation for $k^{-}$.

Generally an equation of the form
\begin{equation}
z^{4n+1}+z^{4n}-\mu^{2}=0
\end{equation}
with $\mu^{2}=M^{2}/\Lambda^{2}$ has $4n+1$ complex solutions, which in the
following will be denoted as $z_{i}$ where $i=1,\ldots,4n+1$. These
solutions depend on the specific value of $\mu^{2}$ and have to be
calculated numerically. Then the poles of the quark propagators in $k^{-}$
can be found from the following relations:
\begin{align}
\frac{k^{2}}{\Lambda^{2}}-1+i\epsilon & =z_{i},  \label{poles1} \\
\frac{(P-k)^{2}}{\Lambda^{2}}-1+i\epsilon & =z_{i}.  \label{poles2}
\end{align}
We have shown in Ref.\cite{Praszalowicz:2001wy} that all the poles
corresponding to Eq.(\ref{poles1}) should lie below the integration contour
whereas the poles corresponding to Eq.(\ref{poles2}) above. As a result the $%
dk^{-}$ integrals yield real wave functions, which vanish for $u$
outside the region $0<u<1$. Moreover for
$\Lambda\rightarrow\infty$, i.e. for a constant $M(k)$, this
prescription reduces in a continuous way to the standard one of
Feynman.

With this prescription the calculations are rather straightforward and we
obtain:
\begin{equation}
\psi_{\pi}^{AV}(u,k_{T}^{2})=\frac{1}{\Lambda^{2}}\frac{N_{c}M^{2}}{(2\pi
)^{2}F_{\pi}^{2}}\sum\limits_{i,k=1}^{4n+1}f_{i}f_{k}\,\frac{%
z_{i}^{n}z_{k}^{3n}u+z_{i}^{3n}z_{k}^{n}(1-u)}{t+1+z_{i}u+z_{k}(1-u)},
\label{AVukt}
\end{equation}
\begin{equation}
\psi_{\pi}^{PS}(u\,,k_{T}^{2})=\frac{N_{c}M}{(2\pi)^{2}\left\langle \bar {q}%
q\right\rangle }\sum\limits_{i,k=1}^{4n+1}f_{i}f_{k}\frac{%
z_{i}^{3n}z_{k}^{3n}(1+\frac{z_{i}+z_{k}}{2})-\mu^{2}z_{i}^{n}z_{k}^{n}}{%
t+1+uz_{i}+(1-u)z_{k}}  \label{PSukt}
\end{equation}
(where $t=k_{T}^{2}/\Lambda^{2}$). Factors $f_{i}$
\begin{equation}
f_{i}=\prod\limits_{_{\scriptstyle k\neq i}^{\scriptstyle k=1}}^{4n+1}\frac {%
1}{z_{i}-z_{k}}
\end{equation}
obey the following properties
\begin{equation}
\sum\limits_{i=1}^{4n+1}z_{i}^{m}f_{i}=\left\{
\begin{array}{ccc}
0 & \text{for} & m<4n \\
&  &  \\
1 & \text{for} & m=4n
\end{array}
\right.  \label{fiprop}
\end{equation}
which are crucial for the convergence of the $dt$ integrals. The $dt$
integration gives:
\begin{equation}
\phi_{\pi}^{AV}(u)=-\frac{N_{c}M^{2}}{(2\pi)^{2}F_{\pi}^{2}}\sum
\limits_{i,k}f_{i}f_{k}\,\left[ z_{i}^{n}z_{k}^{3n}u+z_{i}^{3n}z_{k}^{n}(1-u)%
\right] \,\ln\left( 1+z_{i}u+z_{k}(1-u)\right)  \label{AVu}
\end{equation}
and
\begin{equation}
\phi_{\pi}^{PS}(u)=-\frac{N_{c}M\Lambda^{2}}{(2\pi)^{2}\left\langle \bar {q}%
q\right\rangle }\sum\limits_{i,k}f_{i}f_{k}\left[ z_{i}^{3n}z_{k}^{3n}(1+%
\frac{z_{i}+z_{k}}{2})-\mu^{2}z_{i}^{n}z_{k}^{n}\right] \ln
(1+uz_{i}+(1-u)z_{k}).  \label{PSu}
\end{equation}
The normalization condition for $\phi_{\pi}^{AV}$ is used to fix the value
of the cutoff parameter $\Lambda$ for given $n$ with $F_{\pi}$ fixed at the
physical value of 93 MeV. Then the normalization condition for $\phi_{\pi
}^{PS}$ can be used to calculate the quark condensate.

\section{Gluon condensate}

\label{s:GG}

The Euclidean formula for the gluon condensate in the instanton model of the
QCD vacuum reads \cite{DPrev,DP}: 
\begin{equation}
\left\langle \frac{\alpha }{\pi }GG\right\rangle =32N_{c}\int \frac{%
d^{4}k_{E}}{(2\pi )^{4}}\frac{M^{2}(k_{E})}{k_{E}^{2}+M^{2}(k_{E})}.
\label{GGcond}
\end{equation}
In the instanton model, assuming that the gluon condensate is known, Eq.(\ref
{GGcond}) should be understood as a gap equation for the value $M=M(0)$,
since the shape of the $k$ dependence is uniquely determined as a Fourier
transform of a fermion zero mode in the presence of the instanton
configuration. Alternatively, for given $M$, one can use Eq.(\ref{GGcond})
to calculte the gluon condensate.

A comparison of equations (\ref{qbarqE}) and (\ref{GGcond}) for a constant
constituent mass $M$ (assuming some Euclidean cutoff) suggests the following
relation between the quark and gluon condensates: 
\begin{equation}
{\cal M}\left\langle \bar{q}q\right\rangle =-c\left\langle \frac{\alpha }{%
\pi }GG\right\rangle .  \label{qbqGG8}
\end{equation}
Here ${\cal M}=M$ (constituent quark mass) and $c=1/8$. Let us remind that
relation (\ref{qbqGG8}) follows from the instanton model of the QCD vacuum
in the zero mode approximation and in the chiral limit ($m=0$) \cite{DP}. It
would have to be substantially modified for heavy current quark mass $m$ 
\cite{Pob,Mus}. In
Ref. \cite{DubSmil} a similar relation has been obtained for an arbitrary
self-dual gluonic field (in a {\em dilute} gas approximation) with, however, 
${\cal M}=$ $m$ (current quark mass of a light quark). In contrast to the
previous case such a relation is not satisfied phenomenologically and
moreover it cannot be true in the chiral limit. On the other hand an
expansion for {\em large} current quark mass $m_{q}$ \cite{NSVZ} yields in
the leading order relation (\ref{qbqGG8}) with ${\cal M}=$ $m_{q}$ and $%
c=1/12$ \cite{ERT}. It has been argued in Ref.\cite{ERT} that this relation
holds also for ${\cal M}=M$ (constituent quark mass). Whether this
identification can be theoretically justified is not obvious, since $M$
being of the order of 300~MeV is probably not large enough to be considered
as heavy. It is, however, beyond the scope of this paper to discuss here the
differences between the approaches of Refs.\cite{DubSmil}, \cite{ERT} and 
\cite{DP} (see, however \cite{BroDo}). We simply use Eq.(\ref{GGcond}) as
derived in Ref.\cite{DP} and calculate the gluon condensate.

The first estimate of the gluon
condensate given already in Ref.\cite{SVZ}, $(330$~MeV$)^4$, is nowadays
believed to be too small. In fact different authors advocated different values
for this quantity (see {\em e.g.} Fig.4 of Ref.\cite{Steele} and references
therein). For the purpose of this work we adopt more recent estimate
from Refs.\cite{Nar}
\begin{equation}
\left\langle \frac{\alpha}{\pi}GG\right\rangle=
(393^{+29}_{-38} \;\;{\rm MeV})^4.
\label{GGphen}
\end{equation}
In our case we shall simply use Eq.(\ref{GGcond}) to calculate $%
\left\langle \frac{\alpha}{\pi}GG\right\rangle $ and compare with
Eq.(\ref{GGphen}).

It is tempting to use our experience from the wave function calculation to
continue Eq.(\ref{GGcond}) to the Minkowski space. Indeed equations (\ref
{GGcond}) and (\ref{qbarqE}) differ only by one power of $\sqrt {M(k)M(k-P)}$
(apart from trivial numerical factors). This suggests the following
generalization of (\ref{GGcond}):
\begin{equation}
\left\langle \frac{\alpha}{\pi}GG\right\rangle =i32N_{c}\int\frac{d^{4}k}{%
(2\pi)^{4}}M(k)M(k-P)\,\frac{k(k-P)-M(k)M(k-P)}{%
(k^{2}-M^{2}(k))((k-P)^{2}-M^{2}(k-P))}.  \label{GGcondM}
\end{equation}
For a constant $M$ Eq.(\ref{GGcondM}) can be easily obtained from (\ref
{GGcond}) by first continuing to the Minkowski metric and second by shifting
the integration variable by a constant fourvector $P$ (with $P^{2}=0$):
\begin{align*}
\left\langle \frac{\alpha}{\pi}GG\right\rangle & =i32N_{c}\int\frac{d^{4}k}{%
(2\pi)^{4}}\frac{M^{2}}{k^{2}-M^{2}} 
 =i16N_{c}\int\frac{d^{4}k}{(2\pi)^{4}}\left( \frac{M^{2}}{k^{2}-M^{2}}+%
\frac{M^{2}}{(k-P)^{2}-M^{2}}\right) \\
& =i32N_{c}\int\frac{d^{4}k}{(2\pi)^{4}}\frac{M^{2}(k(k-P)-M^{2})}{%
(k^{2}-M^{2})((k-P)^{2}-M^{2})}.
\end{align*}
Accepting Eq.(\ref{GGcondM}) as a continuation from the Euclidean to the
Minkowski metric we can calculate the gluon condensate for the cutoff
function of Eq.(\ref{cutform}):
\begin{equation}
\left\langle \frac{\alpha}{\pi}GG\right\rangle =-\frac{8N_{c}M^{2}\Lambda^{2}%
}{(2\pi)^{2}}\int du\,\,dt\sum\limits_{i,k}f_{i}f_{k}\frac{%
z_{i}^{2n}z_{k}^{2n}(1+\frac{z_{i}+z_{k}}{2})-\mu^{2}}{t+1+uz_{i}+(1-u)z_{k}}%
.
\end{equation}
The $dt$ integration is trivial (note that the large $t$ part vanishes due
to the properties of the sums of the $f_{i}$ factors). Also the $du$
integral is straightforward. Finally we obtain
\begin{align}
\left\langle \frac{\alpha}{\pi}GG\right\rangle & =\frac{8N_{c}M^{2}%
\Lambda^{2}}{(2\pi)^{2}}\left( \sum\limits_{i}f_{i}^{\,2}\left[
z_{i}^{4n}(1+z_{i})-\mu^{2}\right] \ln(1+z_{i})\right.  \nonumber \\
& 
+\sum\limits_{i\neq k}f_{i}f_{k}\left[
z_{i}^{2n}z_{k}^{2n}(1+\frac{z_{i}+z_{k}}{2})-\mu^{2}\right]  
\left. \left[ \frac{(1+z_{i})\ln(1+z_{i})-(1+z_{k})\ln(1+z_{k})}{%
z_{i}-z_{k}}-1\right] \right)
\end{align}

\begin{figure}[h]
\begin{center}
\epsfig{file=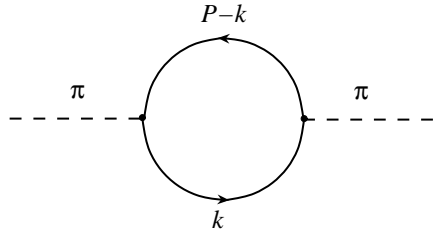,width=6.0cm,clip=}
\end{center}
\caption{Pion self-energy $-i \Sigma_{\protect\pi} (P)$}
\end{figure}
%

It is interesting to observe that the Minkowskian integral
(\ref{GGcondM}) is in fact proportional to the pion self-energy
due to the quark loop calculated in the present effective model.
Indeed
\begin{equation}
\Sigma _{\pi }(P)=-i\frac{2N_{c}}{F_{\pi }^{2}}\int d^{4}k\,M(k)M(k-P)
\text{Tr}\left[\gamma _{5}\frac{1}{\rlap{/}k-M(k)}\gamma _{5}\frac{1}{\left(
\rlap{/}k-\rlap{/}P\right) -M(k-P)}\right]   \label{Sigpi}
\end{equation}
which yields an interesting relation
\begin{equation}
\left\langle \frac{\alpha }{\pi }GG\right\rangle =4F_{\pi }^{2}\Sigma _{\pi
}(P).
\end{equation}
Let us note that in the nonlinear model which we are dealing with, there
exists another loop diagram with two pions coupled to a fermion line at the
same point, which cancels out (\ref{Sigpi}) so that the total pion
self-energy is zero as it should be.

\section{Numerical results}

\label{s:num}

\subsection{Quark and gluon condensates}

Our numerical results are presented in Table \ref{tb:cond}. For 4 different
values of the constituent mass $M=M(0)$ and different $n$ (see Eq.(\ref
{cutform})) we have adjusted the cutoff parameter $\Lambda$ by imposing the
normalization condition on $\phi_{\pi}^{AV}$. In fact, as discussed at
length in Ref.\cite{Praszalowicz:2001wy} the leading twist pion wave
function $\phi_{\pi}^{AV}(u) $ does not change any more if we increase $n$
above $5$. On the other hand for $n\geq5$ the cutoff function (\ref{cutform}%
), if continued to the Euclidean metric, starts to deviate significantly
from the one obtained in the instanton model. Therefore we have chosen to
work with $n_{\text{max}}=5$. From our previous study it seems that the best
agreement with the recent analysis of the CLEO data is obtained for $M=325$
MeV and $n=2-5$ or $M=350$ MeV and $n=2$.

In the last two columns of Table \ref{tb:cond} we have displayed the values
of the quark and gluon condensates. The gluon condensate is quite
insensitive to the parameters of the cutoff function and also to the value
of the constituent quark mass $M$. The numerical value being of the order $%
(380)^{4}$--$(400)^{4}$ MeV$^{4}$ agrees perfectly with the
phenomenological value. In contrast, the quark condensate is
sensitive to the value of $n$ and the numerical value lies between
$-(258)^{3}$ and $-(338)^{3}$ MeV$^{3}$ overshoots the
phenomenological value of $-(250)^{3}$ MeV$^{3}$. The reason for
this
sensitivity is relatively broad $k_{T}^{2}$ distribution of $\tilde{\phi}%
_{\pi}^{PS}(k_{T}^{2})$. We shall come back to this point in Sect.\ref{s:kt}.

\subsection{Pseudo-scalar pion wave function}

The function $\phi_{\pi}^{PS}(u)$ has been calculated within the QCD sum
rules in Refs.\cite{BrauFil2,Ball:1999je}. It had a $u-$ shape and did not
vanish at the end points. In Fig.2 we plot our results for $%
\phi_{\pi}^{PS}(u)$. It can be seen, that for $n=1$ $\phi_{\pi}^{PS}(1)=%
\phi_{\pi}^{PS}(0)\neq0$, whereas for $n\geq2$ $\phi_{\pi}^{PS}(1)=\phi_{%
\pi}^{PS}(0)=0$. The end point behavior is governed by the sum
\begin{equation}
\phi_{\pi}^{PS}(1)\sim\sum\limits_{i}f_{i}\ln(1+uz_{i})\sum\limits_{k}f_{k}
\left[ z_{i}^{3n}z_{k}^{3n}+\frac{1}{2}z_{i}^{3n+1}z_{k}^{3n}+\frac {1}{2}%
z_{i}^{3n}z_{k}^{3n+1}-\mu^{2}z_{i}^{n}z_{k}^{n}\right]  \label{endp}
\end{equation}
which vanishes due to the property (\ref{fiprop}) except for $n=1$ where $%
3n+1=4n$.

\begin{table}[h]
\caption{Quark and gluon condensates for different choices of cutoff
parameters $M(0)$, $n$ and $\Lambda$.}
\label{tb:cond}%
\begin{tabular}{ccccc}
$M(0)$ & $n$ & $\Lambda$ & $-\,\left\langle \bar{q}q\right\rangle $ & $%
\left\langle \frac{\alpha_{s}}{\pi}GG\right\rangle $ \\
{Mev} &  & MeV & {MeV}$^{3}$ & {MeV}$^{4}$ \\ \hline
& 1 & 1249 & $(328)^{3}$ & $(403)^{4}$ \\
325 & 2 & 1862 & $(294)^{3}$ & $(396)^{4}$ \\
& 5 & 3033 & $(280)^{3}$ & $(393)^{4}$ \\ \hline
& 1 & 1156 & $(318)^{3}$ & $(399)^{4}$ \\
350 & 2 & 1727 & $(284)^{3}$ & $(392)^{4}$ \\
& 5 & 2819 & $(271)^{3}$ & $(389)^{4}$ \\ \hline
& 1 & 1081 & $(309)^{3}$ & $(395)^{4}$ \\
375 & 2 & 1621 & $(277)^{3}$ & $(388)^{4}$ \\
& 5 & 2649 & $(264)^{3}$ & $(386)^{4}$ \\ \hline
& 1 & 1020 & $(303)^{3}$ & $(392)^{4}$ \\
400 & 2 & 1543 & $(271)^{3}$ & $(386)^{4}$ \\
& 5 & 2512 & $(258)^{3}$ & $(383)^{4}$%
\end{tabular}
\end{table}

\begin{figure}[h]
\begin{center}
\epsfig{file=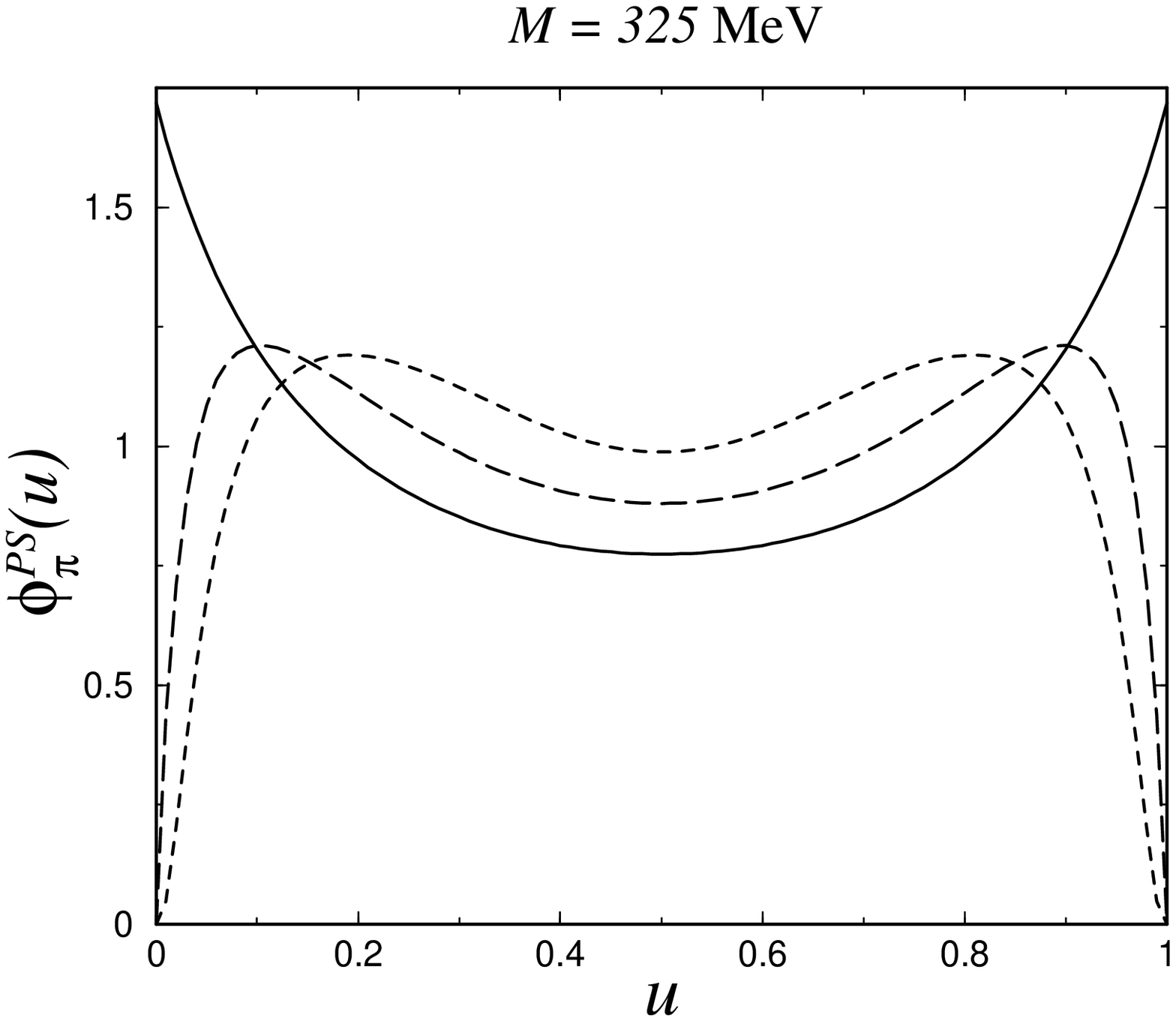,width=6cm}
\epsfig{file=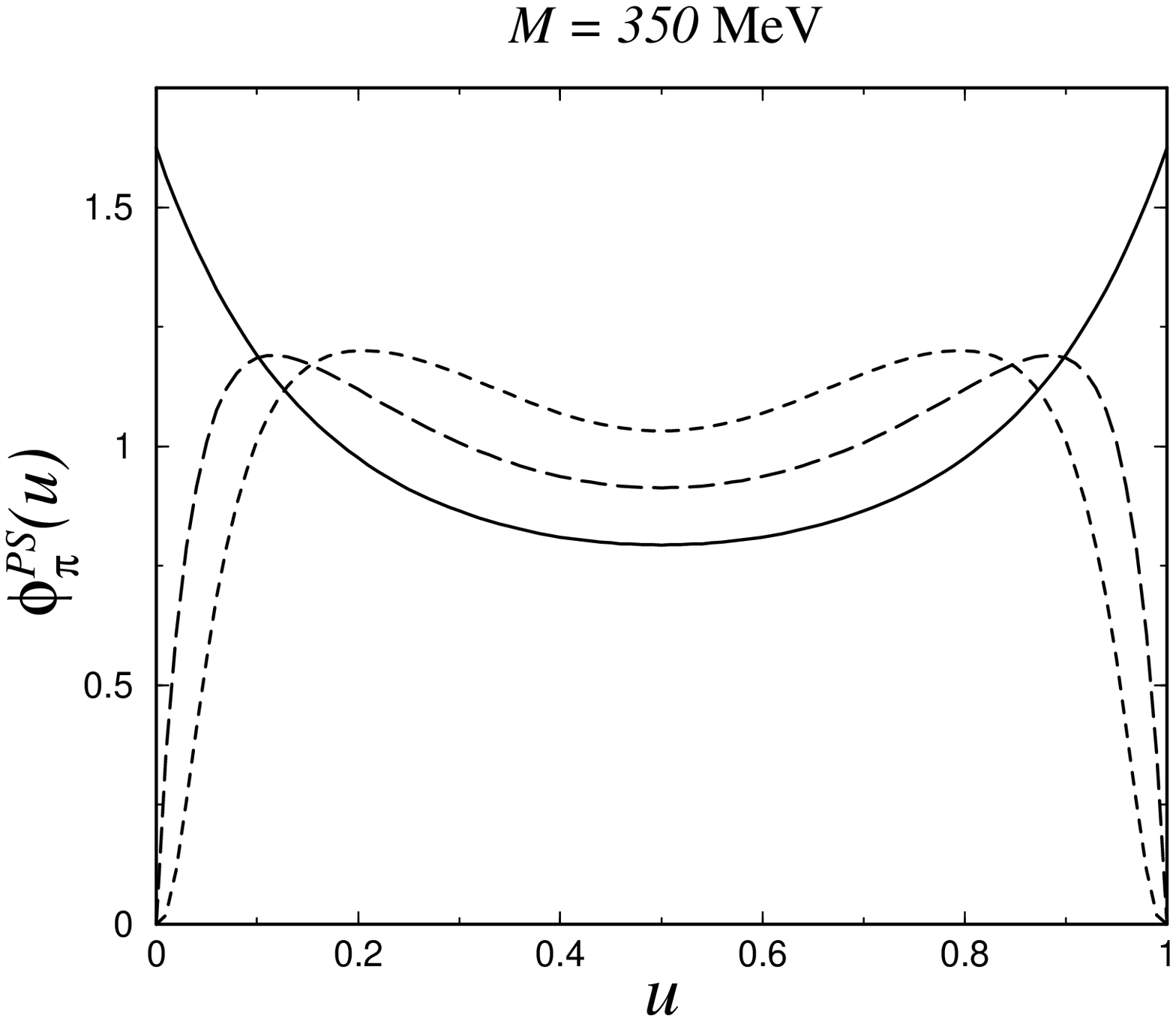,width=6cm} \\
\epsfig{file=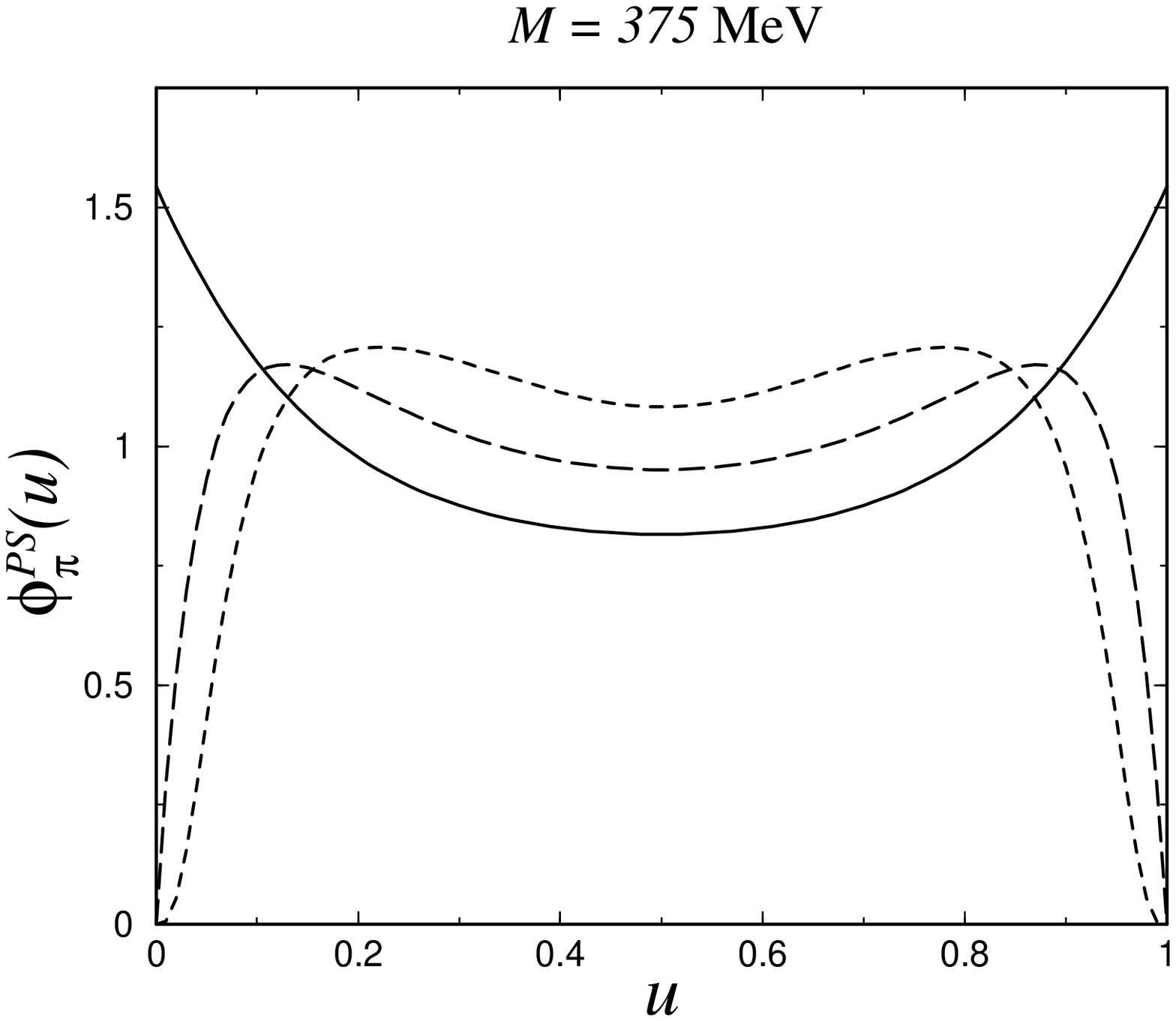,width=6cm}
\epsfig{file=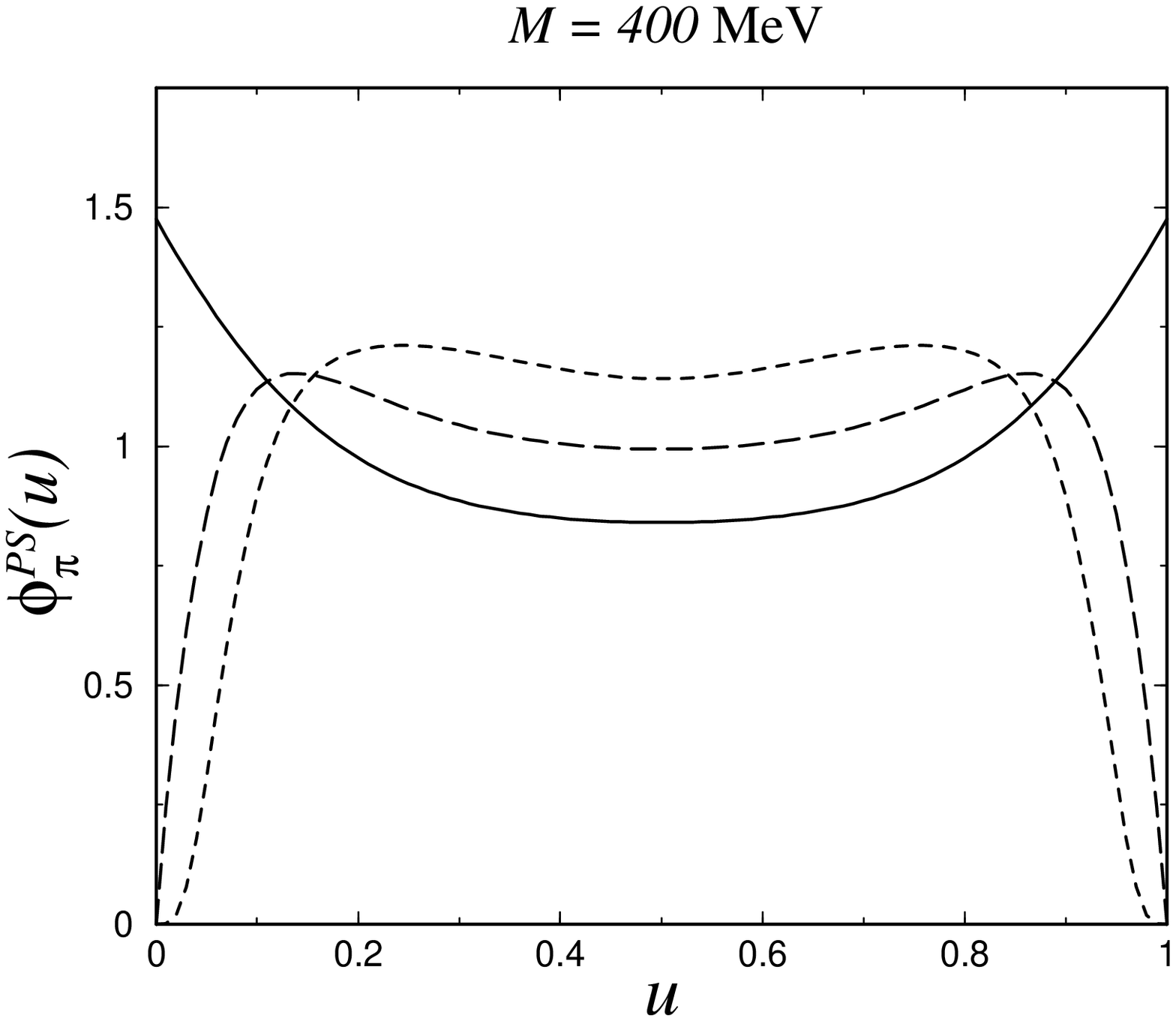,width=6cm}
\end{center}
\caption{Pseudoscalar pion wave function $\protect\phi_{\protect\pi}^{PS}(u)$
for $M=325$, 350, 375 and 400 MeV, for $n=1$ (solid), 2 (long dashed) and 5
(dashed).}
\end{figure}

\subsection{$k_{T}^{2}$ dependence and mixed vacuum condensates}

\label{s:kt}

The large $k_{T}^{2}$ asymptotics of the wave functions and of the
appropriately defined $k_{T}^{2}$ distribution of the gluon condensate
\begin{equation}
\left\langle \frac{\alpha }{\pi }GG\right\rangle =\int\limits_{0}^{\infty
}dk_{T}^{2}\,\tilde{G}(k_{T}^{2})
\end{equation}
is as follows
\begin{equation}
\tilde{\phi}_{\pi }^{AV}(k_{T}^{2})\sim \left( \frac{1}{k_{T}^{2}}\right)
^{4n+1},\qquad \tilde{\phi}_{\pi }^{PS}(k_{T}^{2})\sim \left( \frac{1}{%
k_{T}^{2}}\right) ^{2n},\qquad \tilde{G}(k_{T}^{2})\sim \left( \frac{1}{%
k_{T}^{2}}\right) ^{4n}  \label{ktasymp}
\end{equation}
Therefore for example for $n=1$ already the first moment of $\phi _{\pi
}^{PS}(k_{T}^{2})$ is divergent.

As a kind of remedy for this divergence let us try to introduce a
cutoff
for $dk_{T}^{2}$ integration $T=K_{T}^{2}/\Lambda^{2}$. In Table II we list $%
k_{T}^{2}$ moments, $F_{\pi}$ and $\left\langle \frac{\alpha}{\pi }%
GG\right\rangle $ for $T$ chosen in such a way that the quark
condensate equals $\left\langle \bar{q}q\right\rangle =-(250$
MeV$)^{3}$. Interestingly, by fixing $\left\langle
\bar{q}q\right\rangle $ with one parameter $T$ we are able to
reproduce ratio (\ref{ratio}) with high
accuracy for almost all sets of parameters. Unfortunately the ratio $%
\left\langle k_{T}^{4}\right\rangle _{AV}/\left\langle
k_{T}^{4}\right\rangle _{PS}$ is less stable: $0.27-0.43$. Moreover it is
more than two times smaller than the one obtained by means of Eqs.(\ref{kt4}%
) in Ref.\cite{Zhitkt4}.

One should, however, keep in mind that the absolute values of
$k_{T}^{2}$ moments, especially in the PS channel, depend strongly
on $T$. This is illustrated in Table~\ref{tb:compare} for $M=350$
MeV and $n=2$.

One more remark is here in order. We have not readjusted $\Lambda$ since $%
F_{\pi}(T=0.25)\sim92$ MeV instead of $93$, a negligible change. In Fig.4
we plot $T$ dependence of the ratios $F_{\pi}(T)/F_{\pi}$, $%
\sqrt[3]{\left\langle \bar{q}q\right\rangle (T)}/\sqrt[3]{\left\langle \bar{q%
}q\right\rangle }$ and $\sqrt[3]{\left\langle \frac{\alpha}{\pi }%
GG\right\rangle (T)}/\sqrt[3]{\left\langle \frac{\alpha}{\pi}GG\right\rangle
}$ for $n=2$. We see that while $F_{\pi}$ and gluon condensate saturate
relatively fast, the quark condensate (that is $\int dk_{T}^{2}\tilde{\phi }%
_{\pi}^{PS}(k_{T}^{2})$) saturates very slowly.

\begin{figure}[h]
\begin{center}
\epsfig{file=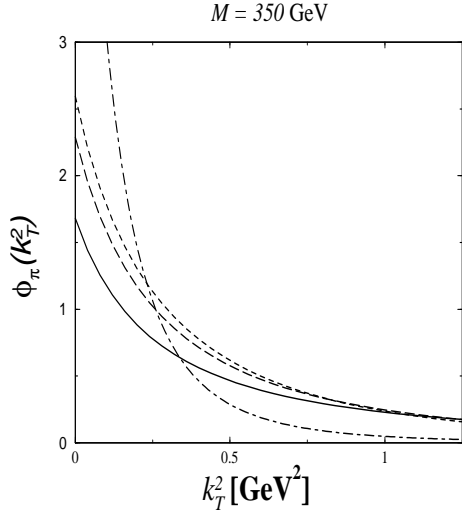,width=6cm,height=7cm,clip=}
\end{center}
\caption{Dependence on $k_{T}^{2}$ of $\protect\phi_{\protect\pi}^{PS}(u)$
for $M=350$~MeV, for $n=1$ (solid), 2 (long dashed) and 5 (dashed) and
of $\protect\phi_{\protect\pi }^{AV}(u)$ for all $n=1 \ldots5$ (dot-dashed)
}
\end{figure}

\begin{table}[tbp]
\caption{$F_{\protect\pi}$, gluon condensate and $k_T^2$ moments of
axial-vector and pseudo-scalar pion wave functions with transverse
integration cutoff fixed by requirement that $\left\langle \bar{q}%
q\right\rangle =-(250$ MeV$)^{3}$.}
\label{tb:zcutoffem}
\begin{center}
\begin{tabular}{lllrcccccccc}
$M$ & $n$ & $T$ & $K_T$& $F_{\pi}$ & $\sqrt[3]{\left\langle \frac{\alpha}{\pi }%
GG\right\rangle }$ & \multicolumn{2}{c}{$\sqrt{\left\langle
k_{T}^{2}\right\rangle }$} & $\frac{\left\langle k_{T}^{2}\right\rangle _{AV}%
}{\left\langle k_{T}^{2}\right\rangle _{PS}}$ & \multicolumn{2}{c}{$\sqrt[4]{%
\left\langle k_{T}^{4}\right\rangle }$} & $\frac{\left\langle
k_{T}^{4}\right\rangle _{AV}}{\left\langle k_{T}^{4}\right\rangle _{PS}}$ \\
MeV &  & & MeV & MeV & MeV & \multicolumn{2}{c}{MeV} &  & \multicolumn{2}{c}{MeV}
&  \\
&  &  &  & & & AV & PS &  & AV & PS &  \\ \hline
& 1 & 0.374 & 764 & 89.5 & 378 & 356 & 476 & 0.56 & 428 & 529 & 0.43 \\
325 & 2 & 0.191& 813 & 90.7 & 381 & 370 & 495 & 0.56 & 445 & 554 & 0.42 \\
& 5 & 0.079 & 852 & 91.4 & 382 & 378 & 508 & 0.56 & 456 & 570 & 0.41 \\ \hline
& 1 & 0.47 & 792 & 90 & 380 & 363 & 487 & 0.55 & 436 & 544 & 0.41 \\
350 & 2 & 0.25 & 864& 91.6 & 382 & 378 & 513 & 0.54 & 455 & 577 & 0.39 \\
& 5 & 0.105 &913 & 92 & 383 & 385 & 527 & 0.53 & 465 & 597 & 0.37 \\ \hline
& 1 & 0.58 &823& 91 & 380 & 368 & 499 & 0.54 & 442 & 558 & 0.39 \\
375 & 2 & 0.31 &902& 92 & 382 & 381 & 524 & 0.53 & 459 & 592 & 0.36 \\
& 5 & 0.14&991 & 92.6 & 383 & 388 & 548 & 0.50 & 471 & 626 & 0.32 \\ \hline
& 1 & 0.69 &847 & 91.5 & 380 & 371 & 507 & 0.53 & 445 & 569 & 0.37 \\
400 & 2 & 0.39 &964 & 92.5 & 382 & 383 & 540 & 0.50 & 464 & 615 & 0.32 \\
& 5 & 0.189 &1092 & 92.9 & 382 & 389 & 569 & 0.47 & 474 & 659 & 0.27
\end{tabular}
\end{center}
\end{table}
\begin{table}[tbp]
\caption{Same as in Table \ref{tb:zcutoffem} for $M=350$ MeV and $n=2$
compared with the results with no $k_{T}^{2}$ cutoff.}
\label{tb:compare}
\begin{center}
\begin{tabular}{llccccccc}
$T$ & $-\sqrt[3]{\left\langle \bar{q}q\right\rangle }$ & $F_{\pi}$ & $\sqrt[3%
]{\left\langle \frac{\alpha}{\pi}GG\right\rangle }$ & \multicolumn{2}{c}{$%
\sqrt{\left\langle k_{T}^{2}\right\rangle }$} & $\frac{\left\langle
k_{T}^{2}\right\rangle _{AV}}{\left\langle k_{T}^{2}\right\rangle _{PS}}$ &
\multicolumn{2}{c}{$\sqrt[4]{\left\langle k_{T}^{4}\right\rangle }$} \\
& MeV & MeV & MeV & \multicolumn{2}{c}{MeV} &  & \multicolumn{2}{c}{MeV} \\
&  &  &  & AV & PS &  & AV & PS \\ \hline
0.25 & 250 & 91.6 & 382 & 378 & 513 & 0.54 & 455 & 577 \\
$\infty$ & 284 & 93 & 392 & 420 & 921 & 0.21 & 540 & 1360
\end{tabular}
\end{center}
\end{table}

\begin{figure}[h]
\begin{center}
\epsfig{file=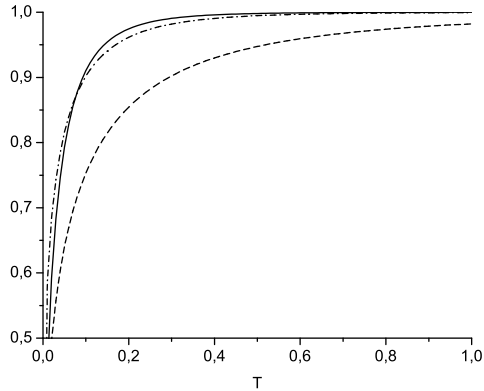,width=7cm}
\end{center}
\caption{Dependence on the cutoff $T$ of the following ratios:
$F_{\protect\pi}(T)/F_{\protect\pi}$
(solid), $\sqrt[3]{\left\langle \bar{q}q\right\rangle (T)}/\sqrt[3]{%
\left\langle \bar{q}q\right\rangle }$ (dashed) and $\sqrt[3]{\left\langle
\frac{\protect\alpha}{\protect\pi}GG\right\rangle (T)}/\sqrt[3]{\left\langle
\frac{\protect\alpha}{\protect\pi }GG\right\rangle }$(dash-dotted) for $n=2$%
. }
\end{figure}

In order to estimate the value of the mixed condensate of dimension 5, $%
\left\langle ig\,\bar{q}\,\sigma \cdot G\,q\right\rangle $, we choose to use
the first equation of Eqs.(\ref{kt2}). Although this equation is only an
approximate one, in the case of no transverse momentum cutoff it gives
relatively stable results due to the fast convergence of the $k_{T}^{2}$
integrals. In the case of the finite transverse momentum cutoff any of the
two equations (\ref{kt2}) can be used, since the relation (\ref{ratio}) is
fulfilled with good accuracy. In the case of no cutoff ({\em i.e.} $T=\infty
$) we get that (for $\left\langle \bar{q}q\right\rangle $ from Table \ref
{tb:cond}):
\begin{equation}
\left\langle ig\,\bar{q}\,\sigma \cdot G\,q\right\rangle =-\,450^{5}\quad
\text{to}\quad -550^{5}\;\text{MeV}^{5}.  \label{C5nocut}
\end{equation}
The value of (\ref{C5nocut}) decreases with increasing $M$ and with
increasing $n$. Interestingly, for the parameters which are closest to the
original instanton model, $M=350$ MeV and $n=2$, we get $-(493$ MeV$)^{5}$
in perfect agreement with the direct calculation of $\left\langle ig\,\bar{q}%
\,\sigma \cdot G\,q\right\rangle $ in the instanton model \cite
{Polyakov:1996kh}, which gives $-(490$ MeV$)^{5}$. In the case of the finite
cutoff the value of $\left\langle ig\,\bar{q}\,\sigma \cdot
G\,q\right\rangle $ is smaller and almost independent of the specific set of
parameters (\ref{cutform}):
\begin{equation}
\left\langle ig\,\bar{q}\,\sigma \cdot G\,q\right\rangle =-\,427^{5}\quad
\text{to}\quad -443^{5}\;\text{MeV}^{5}.  \label{C5cut}
\end{equation}
This range of values is compatible with the result obtained within
the QCD sum rules \cite{BelIof} at low normalization point:
$-(416~{\rm MeV})^5$.

\section{Summary and conclusions}

\label{s:sum}

In this paper, following our previous work \cite{Praszalowicz:2001wy} and
Refs.\cite{PetPob,Bochum}, we have studied a simple, nonperturbative model
in which quarks acquire a momentum dependent dynamical mass (\ref{cutform}).
In Ref.\cite{Praszalowicz:2001wy} we have employed this model to calculate
the axial-vector leading twist pion light cone wave function, whereas here
we have extended our calculations to the case of the pseudo-scalar, twist 3
wave function $\phi _{\pi }^{PS}$, which is normalized to the quark
condensate. Since the model parameters are fixed by the normalization of the
axial-vector wave function we could use the normalization condition for $%
\phi _{\pi }^{PS}$ to calculate the quark condensate. One of the advantages
of the present model is that one can calculate not only the longitudinal
quark distribution in the pion, but also $k_{T}^{2}$ distributions both in
the AV and PS channels. Therefore we could use the relations between $%
k_{T}^{2}$ moments and mixed quark gluon condensates \cite
{ChZhitZhit,Zhitkt4} to estimate the latter in the present model.

One remark is here in order. The advantage of our approach is its
simplicity, both conceptual and technical, which makes it possible
to compute many quantities analytically. One might therefore
suspect that the model is oversimplified and that our assumptions
are too crude to be in general true. For example the model has no
trace of confinement. It is therefore of importance to perform
various tests in order to gain confidence in the model as well as
to find its limitations. This study shows that in fact the model
works much better than one might have initially expected. As shown
in Ref.\cite{Praszalowicz:2001wy} $\phi _{\pi }^{AV}$ is
compatible with recent CLEO data \cite{YakSch}; quark condensate
calculated here from the normalization condition of $\phi _{\pi
}^{PS}$ comes out reasonable, although bigger than the
phenomenological value. The gluon condensate, which we have
computed from the gap equation (\ref{GGcond}) guessing the
continuation to the Minkowski space, comes out very well. The
mixed
quark-gluon condensate of dimension 5 calculated from the value of $%
\left\langle k_{T}^{2}\right\rangle _{AV}$ (see Eq.(\ref{kt2})) is in a
surprising agreement with the direct evaluation of $\left\langle ig\,\bar{q}%
\,\sigma \cdot G\,q\right\rangle $ in the instanton model \cite
{Polyakov:1996kh}.

An obvious limitation of the model is due to the $k_{T}^{2}$ asymptotics of $%
\tilde{\phi}_{\pi }^{PS}$ and also $\tilde{\phi}_{\pi }^{AV}$ (\ref{ktasymp}%
). This powerlike behavior makes higher $k_{T}^{2}$ moments divergent, and
therefore higher mixed condensates uncalculable. For $\tilde{\phi}_{\pi
}^{AV}$ the power is still relatively high, whereas for $\tilde{\phi}_{\pi
}^{PS}$ it is rather low. This is the reason why the quark condensate comes
out higher that the phenomenological value and depends (within a factor of
2) on model parameters. Since there exists another wave function normalized
to $\left\langle \bar{q}q\right\rangle $, namely $\phi _{\pi }^{\sigma }$
\cite{Ball:1999je}, it would be interesting to calculate $\left\langle \bar{q%
}q\right\rangle $ from $\phi _{\pi }^{\sigma }$ and compare with
the present results \cite{Dubr}. As a remedy for the broadness of
$\tilde{\phi}_{\pi }^{PS}(k_{T}^{2})$ we have employed the
$k_{T}^{2}$ cutoff, which does not affect much the gluon condensate
and $F_{\pi }$ ({\em i.e.} $\phi _{\pi }^{AV}$) but allows to
reproduce the phenomenological value of $\left\langle
\bar{q}q\right\rangle =-(250$ MeV$)^{3}$. Then the ratio
$\left\langle k_{T}^{2}\right\rangle _{AV}/\left\langle
k_{T}^{2}\right\rangle _{PS}\sim 5/9$ is in agreement with the
analysis of Ref.\cite{ChZhitZhit}. Unfortunately one single cutoff
is probably not enough to make all higher moments reliable.

Both quark and gluon condensate depend on the scale $Q_0^{2}$. In
model calculations it is, however, hard to define precisely what
value should be taken for $Q_0$. In the instanton model of the QCD
vacuum the scale is determined by the average inverse size of the
instanton $1/\overline{\rho} = 600$~MeV. It is therefore natural
to expect that also in our case the normalization scale is of the
same order. Its connection to the scale $\Lambda$ entering
Eq.(\ref{cutform}) is, however, by no means straightforward since
in fact we use various values of $\Lambda$ and $n$ to approximate
{\em the same} function $M(k)$ evaluated in the instanton model
for $1/\overline{\rho} = 600$~MeV. One way to determine $Q_0^2$ in
the present approach is to identify $Q_0=K_T$, where $K_T$ is the
transverse cutoff displayed in Table~\ref{tb:zcutoffem}. Indeed,
this is precisely the way how the normalization scale is defined
for the light cone wave functions (see {\em e.g.}
Ref.\cite{BrLep}). Numerically it comes out that $ 760 < Q_0 <1100
$ MeV.

The twist 3 pion wave function $\phi _{\pi }^{PS}$, which we have
plotted in Fig.2 is theoretically an interesting object, since it
tends asymptotically to 1, and therefore might be enhanced in
certain kinematical regions. The calculations of $\phi _{\pi
}^{PS}$ from the QCD sum rules suggest that it does not vanish at
the end points \cite{BrauFil2,Ball:1999je}. We find this
kind of behavior only for $n=1$ (see Eq.(\ref{cutform})), whereas for $n>1$ $%
\phi _{\pi }^{PS}$ vanishes at the end points. Interestingly, the
vanishing of $\phi _{\pi }^{PS}$ for $u=0,1$ is correlated with
the nonconvexity of the axial-vector wave function at the end
points, which for $u\rightarrow 0$ (or $1$) behaves like $u^{n}$
(or $(1-u)^{n}$). It would be worthwhile to check if this
correlation holds also in other models. The fact, that the
axial-vector pion distribution amplitude may be concave at the end
points has been already pointed out in Ref.\cite{NiSKim} and
confirmed by model calculations within the framework of the
nonlocal sum rules \cite{BaMiSte}.

\vspace{0.3cm}

This work was partially supported by the Polish KBN Grants 
PB~2~P03B~{\-} 019~17 and PB 2 PO3B 048 22. 
M.P. is grateful to W.Broniowski, A.E. Dorokhov, K.Goeke, H.-Ch. Kim,
M.V.Polyakov and N.G.Stefanis for discussions and interesting suggestions.

\end{document}